\documentstyle[aps,epsfig]{revtex}
\begin{document}
\bibliographystyle{physics}
\begin{titlepage}
\begin{center}
{\large New interpretation to zitterbewegung}

\footnote{Email address: wangzhiyong188@163.net}
\footnote{Email address: zhangal@itp.ac.cn}
\vspace{0.4cm}
{\small Zhi-Yong Wang$^{a,b}$
and Ailin Zhang$^{c}$}\\
\vspace{0.4cm}
{\small $^a$ P. O. Box 1, Xindu, Chengdu, Sichuan, 610500, P. R. China}\\
{\small $^b$ University of Electronic Science and Technology of China,
Chengdu, 610054, P. R. China}\\
{\small $^c$ Institute of Theoretical Physics, Academia Sinica, P. O. Box
2735, Beijing 100080, P. R. China}\\
\end{center}
\vspace{1.0cm}
\begin{abstract}
In previous investigations on zitterbewegung(zbw) of electron, it is
believed that the zbw results from some internal motion of electron.
However, all the analyses are made at relativistic quantum mechanical
level. In framework of quantum field theory (QFT), we find that the origin
of zbw is different from previous conclusion. Especially, some new
interesting conclusions are derived at this level: 1) the zbw arises from
the rapid to-and-fro polarization of
the vacuum in the range of the Compton wavelength (divided by $4\pi$) of
the electron, which offer the four-dimensional(4D) spin and intrinsic
electromagnetic-moment tensor to the electron; 2) Any attempt that
attributes spin (rather than double the spin) of the electron to some kind
of orbital angular momentum would not be successful; 3) the macroscopic
classical speed of the Dirac vacuum medium vanish in all inertial systems.

\vskip 0.5 true cmPACS~ Indices: 12.20.-m, 11.40.-q, 03.50.-z, 11.90.+t
\end{abstract}
\vspace{2cm}\vfill
\end{titlepage}
\section{Introduction}

The investigation of phenomenon of zitterbewegung(zbw) has made great
process since it was first proposed by Schr$\ddot
o$dinger~\cite{schrodinger}:
the description and interpretation of the
zbw~\cite{huang,braun,rupp,szymanowski,turek,barut}
are tried; many "classical models"
of the Dirac electron basing on the 
zbw~\cite{barut,hestenes,rodrigues,recami,salesi} are constructed; The zbw
of other particles such as neutrino,
quark, photon and Anyon~\cite{pernice,stefano,dalmazi,kobe,klishevich}
are studied; the theories such as black hole, superstring, cosmology,
and gravity, have been constructed by means of the
zbw~\cite{sidharth,sparling,hep,puthoff} also.

As for the origin of the zbw of electron, more and more
people believe that it results from some internal motion of electron.
However, all the work on the origin of zbw is examined only at the
relativistic
quantum mechanical level, where the zbw's interpretations are based on the
probability-amplitude interpretation of wave function. Starting from the
velocity operator or the current density of the electron, people obtained
the zbw term expressed in term of the spinor wave functions and the Dirac
matrices, but they have not so far given any further calculation for this
zbw term, let alone giving its explicit and exact form,
which is exactly the key to a complete and correct understanding of the
zbw.

When we work at the QFT level, we find previous understanding to the
origin of zbw is ambiguous or misleading. In this paper, we study the 
origin of zbw and drew some new interesting conclusions, which would do
help to our understanding to zbw. 

Firstly, starting from the current density of a free electron, we gain a
explicit and exact expression to the zbw current vectors in term of the
creation and annihilation operators as well as the helicity states of
vector
field. We find that the zbw currents come from the convection current and
the current related to the intrinsic electric-moment, while the current
related to the internal magnetic-moment does not contribute to the zbw
current. Furthermore, we draw conclusions as follows: 1) the zbw arises
from the rapid
to-and-fro polarization of the vacuum in the range of the Compton
wavelength(divided by $4\pi$) of the electron, which offers an internal dynamic
degrees of freedom (i.e. the 4D spin), or the 4D internal
electromagnetic-moment tensor,
to the electron. 2) The electron-positron pair arising from the vacuum
polarization forms a vector triplet with the total momentum vanishing,
which
implies that the macroscopic classical speed of the Dirac vacuum medium
vanish
in all inertial systems. 3) The space magnitudes of the zbw are ${m\over
E}{1\over 2E}$
(for the longitudinal component) or ${1\over 2E}$(for the right- and
left-hand circular
components), and the directions of the zbw current are rotating with the
angular velocity $2E$. Then the zbw velocity projections are $\pm{m\over
E}$
(in the longitudinal direction) or $\pm 1$(in the transversal direction).

In addition, we find a united expression for the 4D total
electromagnetic-moment tensor of the electron, which holds in both
classical
and quantum mechanics. In fact, because of the zbw, we can attribute
$2s^{\mu\nu}$
instead of $s^{\mu\nu}$(the 4D-spin tensor of the electron) to the
difference
between the instantaneous and average 4D orbital angular momentum tensor.
In other words, all the traditional attempts~\cite{huang,barut,recami}
that attribute the spin(rather than double the spin) of the electron to
some kind of
orbital angular momentum, would not be successful.

\section{The helicity states of vector field}

The helicity states of vector field play such an important
role that we discuss it in detail firstly. For a vector field, the
operator
of the spin projection in the direction of momentum $\vec{p}$ is
${1\over |\vec{p}|}\vec{\tau}\cdot \vec{p}$, where
\begin{eqnarray}\label{1}
\begin{array}{ccc}
\tau_1=\left (\begin{array}{lcr}0 & 0 & 0 \\
0 & 0 & -i \\
0 & i & 0 \end{array} \right ), &\tau_2=\left (\begin{array}{lcr}0 & 0 & i
\\
0 & 0 & 0 \\  -i & 0 & 0
\end{array} \right ), &\tau_3=\left (\begin{array}{lcr}0 & -i & 0 \\i & 0
& 0 \\
0 & 0 & 0\end{array} \right ).
\end{array}
\end{eqnarray}
are the spin matrices of vector filed. Let
\begin{eqnarray}\label{2}
{1\over |\vec{p}|}\vec{\tau}\cdot \vec{p}
\eta_\gamma=\lambda_\gamma\eta_\gamma,
\end{eqnarray}
we have
\begin{eqnarray}\label{3}
\lambda_1=1,
\eta_{1}={1 \over \sqrt{2}|\vec{p}|} \left (
\begin{array}{lcr}
{p_1p_3-ip_2|\vec{p}| \over p_1-ip_2} \\
{p_2p_3+ip_1|\vec{p}| \over p_1-ip_2} \\
-(p_1+ip_2)
\end{array} \right ), &
\lambda_2=-1,
\eta_2=\eta_1^\ast , &
\lambda_3=0,
\eta_3={1\over |\vec{p}|} \left (
\begin{array}{lcr} 
p_1 \\
p_2 \\
p_3\end{array} \right ).
\end{eqnarray}
where $\eta_1^\ast$ is the complex conjugate of $\eta_1$. Obviously,
$\{\eta_\gamma\}$ form a complete orthonormal basis. On the other hand,
in a coordinate system formed by a orthonormal basis
$\{\vec{e_1},\vec{e_2},\vec{e_3}\}$, where
$\vec{e_3}=\vec{e_1}\times\vec{e_2}={1\over
|\vec{p}|}\vec{p}$,
(\ref{3}) is transformed into the following vector form
\begin{eqnarray}\label{4}
\begin{array}{ll}
\vec{\eta}_1&={1\over \sqrt{2}}(\vec{e_1}+i\vec{e_2})\\
\vec{\eta}_2&={1\over \sqrt{2}}(\vec{e_1}-i\vec{e_2})\\
\vec{\eta}_3&=\vec{e_3}.
\end{array}
\end{eqnarray} 

Clearly, $\vec{p}\cdot\vec{\eta}_1=\vec{p}\cdot\vec{\eta}_2=0$ and
$\vec{\eta}_3\parallel \vec{p}$. That is, $\{\vec{\eta}_1,\vec{\eta}_2,
\vec{\eta}_3\}$ is the spinor representation of the vector basis
$\{\vec{e}_1,\vec{e}_2,\vec{e}_3\}$. Correspondingly, the right-hand and
left-hand circular polarization vectors are denoted by 
$\eta_1$ and $\eta_2$, respectively, while the longitudinal
polarization one by $\eta_3$.

\section{The zbw from the viewpoint of QFT}

In free field's case, the Gordon decomposition of the Dirac current is
\cite{peskin}:
\begin{eqnarray}\label{5}
j^\mu=\bar \psi\gamma^\mu\psi={1\over 2m}[\bar\psi\hat p^\mu\psi-(\hat
p^\mu\bar
\psi)\psi]-{i\over 2m}\hat p_\nu (\bar\psi \Sigma^{\mu\nu}\psi),
\end{eqnarray}
where $\bar\psi$ is the adjoint of the spinor wave operator $\psi$
($\bar\psi=\psi^\dagger\gamma^0$),
$\gamma^\mu$
is the usual Dirac matrices. ${1\over 2}\Sigma^{\mu\nu}={i\over
4}[\gamma^\mu,\gamma^\nu]$
is the 4D spin tensor, corresponds to which, we define $G^{\mu\nu}={e\over
2m}\bar\psi\Sigma^{\mu\nu}\psi$ as the internal electromagnetic moment
density tensor, where $e$ denotes the electron charge. Let
$\varepsilon^{ijk}$
denotes the totally antisymmetric tensor with $\varepsilon^{123}=1
(i,j,k=1,2,3)$, then
\begin{eqnarray}\label{6}
m^i={1\over 2}\varepsilon^{ijk}G_{jk}, n^i=G^{i0}
\end{eqnarray}
are the magnetic moment and the electric moment, respectively.

Traditionally, when we study the conserved Noether's charges corresponding
to the Lorentz
invariance of quantum field, our interest is only focused on the spatial
components of
the 4D angular momentum tensor (i.e. the 3D angular momentum), whereas for
the $0-i(i=1,2,3)$
components we haven't so far given any attention to them. However, the
electron possesses
not only an internal magnetic moment but also an internal electric
moment. That is to say,
the relevant components of 4D spin do have observable effects. As for the
components of 4D
orbital angular momentum tensor, we have to be confronted with the notion
of time operator,
which will be discussed in our later paper.

By using (\ref{5}), the spatial components of $j^\mu$ can be written in
form\cite{huang}:
\begin{eqnarray}\label{7}
\vec{j}=\psi^\dagger\vec{\alpha}\psi={1\over 2m}
[\bar\psi\vec{\hat{p}}\psi-(\vec{\hat{p}}\bar\psi)\psi]
+{1\over 2m}\nabla\times(\bar\psi\vec{\Sigma}\psi)-{i\over
2m}{\partial\over \partial t}
(\bar\psi \vec{\alpha}\psi),
\end{eqnarray}
where
\begin{eqnarray}\label{8}
\begin{array}{cc}
\vec{\Sigma}=\left (
\begin{array}{lcr}
\vec{\sigma} & 0\\
0 & \vec{\sigma}
\end{array} \right ), &
\vec{\alpha}=\left (
\begin{array}{lcr}
0 & \vec{\sigma} \\
\vec{\sigma} & 0
\end{array} \right )
\end{array}
\end{eqnarray}
with $\vec{\sigma}$ being the Pauli matrices. Clearly,
\begin{eqnarray}\label{9}
\vec{m}={1\over 2m}\bar\psi\vec{\Sigma}\psi,
\vec{n}=-{i\over 2m}\bar\psi\vec{\alpha}\psi.
\end{eqnarray}
That is, the first term of (\ref{7}) is the so-called convection current,
while the remaining two terms are the contributions coming
from the magnetic moment $\vec{m}$ and the electric moment $\vec{n}$,
respectively.

We write $\psi$ as
\begin{eqnarray}\label{10}
\psi=\sum\limits_{\vec{p},s}\sqrt{{m\over VE}}[c(p,s)u(p,s)e^{-ipx}+
d^\dagger(p,s)v(p,s)e^{ipx}],
\end{eqnarray}
where $s=1,2$, correspond to the spin $\pm{1\over 2}$, respectively.

Let
\begin{eqnarray}\label{11}
c(\vec{p},s)\equiv c(p_0,\vec{p},s)=c(p,s), c(-\vec{p},s)\equiv
c(p_0,-\vec{p},s), etc,
\end{eqnarray}
each term of (\ref{7}) is integrated to yield 
\begin{eqnarray}\label{12}
\int d^3x{1\over 2m}[\bar\psi\hat p\psi-(\hat
p\bar\psi)\psi]=\sum\limits_{\vec{p},s}
{\vec{p}\over E}[c^\dagger(p,s)c(p,s)-d^\dagger(p,s)d(p,s)]\\\nonumber
+\sum\limits_{\vec{p},s}({m\over E}-{E\over m})\lambda_s\vec{\eta}_3
[c^\dagger(\vec{p},s)d^\dagger(-\vec{p},s)e^{i2Et}-c(-\vec{p},s)d(\vec{p},s)
e^{-i2Et}],
\end{eqnarray}
\begin{eqnarray}\label{13}
\int d^3x{1\over 2m}\nabla\times(\bar\psi\vec{\Sigma}\psi)=0,
\end{eqnarray}
\begin{eqnarray}\label{14}
\int d^3x[-{i\over 2m}{\partial \over \partial
t}(\bar\psi\vec{\alpha}\psi)]=
\sum\limits_{\vec{p},s}{E\over
m}\lambda_s\vec{\eta}_3[c^\dagger(\vec{p},s)
d^\dagger(-\vec{p},s)e^{i2Et}-c(-\vec{p},s)d(\vec{p},s)e^{-i2Et}]\\\nonumber
+\sum\limits_{\vec{p}}\sum\limits_{s\neq s'}\sqrt{2}\vec{\eta}_{s'}
[c^\dagger(\vec{p},s)d^\dagger(-\vec{p},s')e^{i2Et}-c(-\vec{p},s')d(\vec{p},s)
e^{-i2Et}],
\end{eqnarray}
where $s,s'=1,2, \lambda_1=1,\lambda_2=-1$ and
$\vec{\eta}_1,\vec{\eta}_2, \vec{\eta}_3$ are given by (\ref{3}).

It can be found from (\ref{12})-(\ref{14}), the zbw current comes
from the convection current and the current is related to the internal
electric-moment, while the current related to the intrinsic
magnetic-moment
does not contribute to the zbw current. 

That is to say, the traditional conclusion that the Gordon decomposition
of $\vec{j}$ can be regarded as the separation of $\vec{j}$ into the
convection current, originating from the moving charge only, and the
current
associated with the internal magnetic-moment of the electron, is
incorrect!

On the other hand, $\vec{j}=\psi^\dagger\vec{\alpha}\psi$ can be directly
integrated to yield
\begin{eqnarray}\label{15}
\int\vec{j}d^3x=\int d^3x\psi^\dagger\vec{\alpha}\psi=
\vec{v}+\vec{z}_\perp+\vec{z}_\parallel,
\end{eqnarray}
where
\begin{eqnarray}\label{16}
\vec{v}=\sum\limits_{\vec{p},s}{\vec{p}\over
E}[c^\dagger(\vec{p},s)c(\vec{p},s)
-d^\dagger(\vec{p},s)d(\vec{p},s)],
\end{eqnarray}
\begin{eqnarray}\label{17}
\vec{z}_\parallel=\sum\limits_{\vec{p},s}{m\over
E}\lambda_s\vec{\eta}_3[c^\dagger(\vec{p},s)
d^\dagger(-\vec{p},s)e^{i2Et}-c(-\vec{p},s)d(\vec{p},s)e^{-i2Et}],
\end{eqnarray}
\begin{eqnarray}\label{18}
\vec{z}_\perp=\sum\limits_{\vec{p}}\sum\limits_{s\neq
s'}\sqrt{2}\vec{\eta}_{s'}
[c^\dagger(\vec{p},s)d^\dagger(-\vec{p},s')
e^{i2Et}-c(-\vec{p},s')d(\vec{p},s)e^{-i2Et}]\\\nonumber
=\sum\limits_{\vec{p}}\{\sqrt{2}\vec{\eta}_1
[c^\dagger(\vec{p},2)d^\dagger(-\vec{p},1)e^{i2Et}
-c(-\vec{p},1)d(\vec{p},2)e^{-i2Et}]\\\nonumber
+\sqrt{2}\vec{\eta}_2[c^\dagger(\vec{p},1)d^\dagger(-\vec{p},2)e^{i2Et}-
c(-\vec{p},2)d(\vec{p},1)e^{-i2Et}]\}.
\end{eqnarray}
Then (\ref{15}) agrees with $(\ref{12})-(\ref{14})$.
The only difference between (\ref{15}) and (\ref{12})-(\ref{14}) is that
the latter
hold only when $m\neq 0$(which, of course, is a consequence of the fact
that the Gordon decomposition holds only when $m\neq 0$ ).

In correspondence to the particles number operator
$\hat{N}=c^\dagger(p,s)c(p,s)$ and $\hat{N}'=d^\dagger(p,s)d(p,s)$, the
operators
\begin{eqnarray}\label{19}
\hat{O}\equiv c^\dagger(\vec{p},s)d^\dagger(-\vec{p},s'),
\hat{O}'\equiv c(-\vec{p},s')d(\vec{p},s), etc
\end{eqnarray}
in (\ref{12})-(\ref{18})($s,s'=1,2$) can be called as the
vacuum-polarization operator, and $\hat{O}|0\rangle$
or $\langle 0|\hat{O}^\dagger$ are the Dirac vacuum-polarization state,
which can be regarded as the macroscopic-motion carrier of the Dirac
vacuum
medium. In view of the fact that the momentum $\vec{p}$ is arbitrary and
the
total momentum of $\hat{O}|0\rangle$ or $\langle 0|\hat{O}^\dagger$
vanish. the macroscopic average velocity of the Dirac vacuum medium
vanishs
in all inertial systems (note that $\vec{j}$ in (\ref{15}) is free).

$\vec{\eta}_3$ emerging in $\vec{z}_\parallel$ is parallel to $\vec{p}$,
while $\vec{\eta}_1$
and $\vec{\eta}_2$ emerging in $\vec{z}_\perp$ are perpendcular to
$\vec{p}$;
On the other hand, the spin of the states $\vec{z}_\parallel |0\rangle$ or
$\langle 0|\vec{z}_\parallel$ is 0, while the spin of the states
$\vec{z}_\perp |0\rangle$
or $\langle 0|\vec{z}_\perp$ is $\pm 1$. Then, $\vec{z}_\parallel$
corresponds to a longitudinal-polarization vector current (note that $m=0,
\vec{z}_\parallel=0$,
the mass is always related to the longitudinal component of vector); 
and $\vec{z}_\perp$
is a transversal-polarization vector current, which includes the
right-hand (related to $\vec{\eta}_1$) and left-hand (related to
$\vec{\eta}_2$) circular-polarization vector currents.
In other words, the electron-positron pair arising from the vacuum
polarization forms a vector triplet with the total momentum vanishing.

It can be seen from $\int\vec{z}_\parallel dt$ and $\int \vec{z}_\perp
dt$,
the spatial magnitudes of the zbw are ${m\over E}{1\over 2E}$(for the
longitudinal component) or ${1\over 2E}$(for the right- and left-hand
circular components), and the directions of the zbw current are rotating
with the angular velocity $2E$. Then, 
the zbw velocity projections are $\pm{m\over E}$(in the longitudinal
direction)
or $\pm 1$(in the transversal direction).

In summary, the zbw arises from the rapid to-and-fro polarization of the
vacuum with the spatial range of ${m\over E}{1\over 2E}$ or ${1\over 2E}$,
which, just as we will discuss below, offers an internal dynamic degree
of freedom (i.e. the 4D-spin) to the electron. As a result, the zbw motion
is an internal one independent of the macroscopic classic motion.
Accordingly, the zbw current in (\ref{15}) doesn't vanish as $\vec{p}=0$.

\section{The 4D total electromagnetic-moment tensor of the electron}

In view of $x^0\equiv t$ and
$\dot{x}^\mu\equiv {\partial x^\mu\over \partial t}=(1,\vec{\alpha})$,
we have
\begin{eqnarray}\label{20}
\psi^\dagger \dot{x}^\mu \psi=\bar\psi\gamma^\mu\psi=j^\mu.
\end{eqnarray}
Then, the 4D total electromagnetic-moment tensor of the electron can be
defined
as
\begin{eqnarray}\label{21}
M^{\mu\nu}\equiv {e\over 2}(x^\mu \dot{x}^\nu- x^\nu \dot{x}^\mu),
\end{eqnarray}
whose components are the total electric- and magnetic-dipole moment of the
electron. The definition (\ref{21}) holds in both classic and quantum
cases.
It is easy to show that $\gamma^0 m$ and $V^\mu\equiv \gamma^0
\dot{x}^\mu=\gamma^\mu$
are the relativistic mass and 4D-velocity operator of the electron,
respectively.
Due to the zbw, $V^\mu$ is different from the usual 4D-velocity. We define
\begin{eqnarray}\label{22}
p^\mu_{ins}\equiv m V^\mu, L^{\mu\nu}_{ins}\equiv x^\mu p^\nu_{ins}-x^\nu
p^\mu_{ins}
\end{eqnarray}
as the 4D instantaneous momentum and the 4D instantaneous orbital angular
momentum, respectively. Clearly,
\begin{eqnarray}\label{23}
\psi^\dagger M^{\mu\nu}\psi={e\over 2m}\bar\psi(x^\mu p^\nu_{ins}-
x^\nu p^\mu_{ins})\psi={e\over 2m}\bar\psi L^{\mu\nu}_{ins}\psi.
\end{eqnarray}
By applying
$\gamma^\mu\gamma^\nu=g^{\mu\nu}-i\Sigma^{\mu\nu}$(where $g^{\mu\nu}$
is the metric tensor with diag(1, -1, -1, -1)), we have
\begin{eqnarray}\label{24}
\psi^\dagger M^{\mu\nu}\psi={e\over 2}\bar\psi(x^\mu \gamma^\nu-
x^\nu\gamma^\mu)\psi={e\over 2m}\bar\psi(L^{\mu\nu}+2S^{\mu\nu})\psi,
\end{eqnarray}
where $L^{\mu\nu}=x^\mu p^\nu-x^\nu p^\mu$ is the usual 4D momentum
operator,
$S^{\mu\nu}={1\over 2}\Sigma^{\mu\nu}$ is the 4D spin. By using
(\ref{23}) and (\ref{24}), we have
\begin{eqnarray}\label{25}
\bar\psi(2S^{\mu\nu})\psi=\bar\psi(L^{\mu\nu}_{ins}-L^{\mu\nu})\psi.
\end{eqnarray}

That is, $2S^{\mu\nu}$(rather than $S^{\mu\nu}$), the difference between
the
instantaneous and the average orbital angular momentum, has the property
of
orbital angular momentum, which implies that one cannot attribute the 4D
spin
$S^{\mu\nu}$ of the electron to any kind of orbital angular momentum.
On the other hand, it is the zbw that results in the differences between
all above instantaneous quantities and the corresponding average ones.
Then the zbw is the sole origin of the 4D spin of the electron, while the
origin of the zbw lies in the rapid to-and-fro polarization of the vacuum
within the spatial range ${m\over E}{1\over 2E}$ or ${1\over 2E}$ of the
electron.

\end{document}